# A New Weighted Information Generating Function for Discrete Probability Distributions

*Amit Srivastava\* and Shikha Maheshwari\*\**

*Department of Mathematics, Jaypee Institute of Information Technology*
*Noida (Uttar Pradesh), India*
*Email: raj_amit377@yahoo.co.in\*, maheshwari.shikha23@gmail.com\*\**

**Abstract**: *The object of this paper is to introduce a new weighted information generating function whose derivative at point 1 gives some well known measures of information. Some properties and particular cases of the proposed generating function have also been studied.*

**Keywords:** *Information generating function, discrete probability distribution, utility distribution.*

## 1. Introduction

The moment generating function of a probability distribution is a convenient mean of evaluating mean, variance and other moments of a probability distribution and an effective embodiment of properties of the same for various analytical processes. The successive derivatives of the moment generating function at point 0 gives the successive moments of a probability distribution if these moments exist. In a similar way, the successive derivatives of the information generating function(IGF) of a probability distribution evaluated at point 1 gives some statistical entities associated with the probability distribution. The information generating function was introduced by S. Golomb [2] in a correspondence and is given by

(1) $$I(t) = \sum_{i \in N} p_i^{\ t}$$

where $\{p_i\}$ is a complete probability distribution with $i \in N$, $N$ being a discrete sample space and $t$ is a real or a complex variable. The first derivative of the above function, at point $t = 1$, gives the negative Shannon's entropy of the corresponding probability distribution i.e. we have

(2) $$-\left(\frac{\partial I(t)}{\partial t}\right)_{t=1} = -\sum_{i \in N} p_i \log p_i = H_n(P)$$

where $H_n(P)$ is the well known Shannon's entropy[5]. Further we have

(3) $$(-1)^r \left(\frac{\partial^r I(t)}{\partial^r t}\right)_{t=1} = (-1)^r \sum_{i \in N} p_i \, (\log p_i)^r$$

Except for factor of $(-1)^r$, the $r^{th}$ derivative of IGF given by (1) gives the $r^{th}$ moment of the self-information of the distribution.

This technique works equally well for discrete and continuous distributions. Moreover the information generating function summarizes those aspects of the distribution which are invariant under measure preserving rearrangements of the probability space. Golomb obtained simple expressions of the IGF defined by (1) for uniform, geometric and $\beta$ - power distributions. However, the quantity (2) measures the average information associated with probabilities of a number of events but does not take into account the effectiveness or importance of these events. Belis and Guiasu [1] raised the very important issue of integrating the quantitative, objective and probabilistic concept of information with the qualitative, subjective and non–stochastic concept of utility. They laid down the two possible postulates for this purpose viz.

- The 'useful' information from two independent events is the sum of the 'useful' information given by two events separately.
- The 'useful' information given by an event is directly proportional to its utility.

On the basis of these postulates, they proposed the following weighted measure of information

(4) $$-\sum_{i \in N} u_i \, p_i \log p_i = H(P, U)$$

where the utility distribution is $U = (u_1, u_2, \dots, u_n, \dots)$ and the probability distribution is $P = (p_1, p_2, \dots, p_n, \dots)$.

The measure (4) is associated with the following utility information scheme [4]

(5) $$\begin{Bmatrix} E_1 & E_2 & \dots & E_n & \dots \\ p_1 & p_1 & \dots & p_n & \dots \\ u_1 & u_2 & \dots & u_n & \dots \end{Bmatrix}$$

$$0 \leq p_i \leq 1, i \in N, \sum_{i \in N} p_i = 1, u_i > 0$$

Here $(E_1, E_2, \dots, E_n, \dots)$ denote a family of events with respect to some random experiment and $u_i$ denotes the utility of an event $E_i$ with probability $p_i$. In general, the utility $u_i$ of an event is independent of its probability of occurrence $p_i$.

Analogous to (4), Hooda and Bhaker [3] defined the following weighted information generating function

(6) $$M(P,U,t) = \sum_{i \in N} u_i p_i^t, t \geq 1$$

Here also $(p_1, p_2, ..., p_n, ...)$ and $(u_1, u_2, ..., u_n, ...)$ are the probability and utility distributions are respectively as defined in (4) and $t$ is a real or a complex variable. Further, we have

(7) $$-\left(\frac{\partial M(P,U,t)}{\partial t}\right)_{t=1} = -\sum_{i \in N} u_i p_i \log p_i = H(P,U)$$

where $H(P,U)$ is measure given by (4).

In this paper, we have defined a new weighted information generating function whose derivative at point 1 gives measure (4). Some properties and particular cases of this new function have also been discussed. Without essential loss of insight, we have restricted ourselves to discrete probability distributions.

## 2. New Information Generating Function

Consider the following function

(8) $$I(P,U,t) = \sum_{i \in N} p_i^{1-u_i(1-t)}, t \geq 1$$

Here $P = (p_1, p_2, ..., p_n, ...)$ and $U = (u_1, u_2, ..., u_n, ...)$ are the probability and utility distributions respectively as defined in (5) and t is a real or a complex variable. Clearly $I(P,U,1) = 1$ and since $0 \leq p_i \leq 1, i = 1, 2, ..., n$, the function (8) is convergent for all $u_i > 0$. If we take $u_i = 1$ for all $i$, the function (8) reduces to (1).

It further follows from (8) that

(9) $$-\left(\frac{\partial I(P,U,t)}{\partial t}\right)_{t=1} = -\sum_{i \in N} p_i \log p_i^{u_i}$$

$$= -\sum_{i \in N} u_i p_i \log p_i = H(P,U)$$

Therefore the function defined by (8) can be defined as the weighted information generating function for the measure defined by (4).

Further we have

(10) $$(-1)^r \left(\frac{\partial^r I(P,U,t)}{\partial^r t}\right)_{t=1} = (-1)^r \sum_{i \in N} p_i (u_i \log p_i)^r$$

The entity $u_i \log p_i$ can be seen as generalized (or weighted) self information for the utility information scheme given by (5). Therefore, except for factor of $(-1)^r$, the $r^{th}$ derivative of weighted IGF given by (8) gives the $r^{th}$ moment of the generalized self-information for the scheme defined by (5).

### Particular Cases

a) Uniform Probability distribution & Constant Utility Distribution

If we consider $p_i = 1/n, i = 1, 2, ...$ and $u_1 = u_2 = ... = u$ (say) then the weighted IGF given by (8) reduces to

$$I(P,U,t) = \sum_{i \in N} \left(\frac{1}{n}\right)^{1-u(1-t)}, t \geq 1$$

Further we have

$$-\left(\frac{\partial I(P,U,t)}{\partial t}\right)_{t=1} = ln(n)^u = u \ln n$$

This is exactly the Shannon entropy for uniform probability distribution and constant utility distribution.

b) Geometric Probability distribution & Constant Utility Distribution

If we consider $p_i = qp^i, i = 0, 1, 2, ..., \infty$ and $u_1 = u_2 = ... = u$ (say) then the weighted IGF given by (8) reduces to

$$I(P, U, t) = \frac{q^{1-u(1-t)}}{1 - p^{1-u(1-t)}}$$

and as a result

$$-\left(\frac{\partial I(P, U, t)}{\partial t}\right)_{t=1} = -u\left(\frac{p \ln p + q \ln q}{q}\right)$$

This is exactly the Shannon entropy for geometric probability distribution and constant utility distribution.

### c) $\beta$ - power Probability distribution & Constant Utility Distribution

If we consider $p_i = \frac{i^{-\beta}}{\zeta(\beta)}, \zeta(\beta) = \sum_{i \in N} i^{-\beta}$ and $u_1 = u_2 = \ldots = u$ (say) then the weighted IGF given by (8) reduces to

$$I(P, U, t) = \sum_{i \in N} \left(\frac{i^{-\beta}}{\sum_{i \in N} i^{-\beta}}\right)^{1-u(1-t)}$$

and as a result

$$-\left(\frac{\partial I(P,U,t)}{\partial t}\right)_{t=1} = u\left(\ln \zeta(\beta) - \frac{\beta \zeta'(\beta)}{\zeta(\beta)}\right)$$

This is exactly the Shannon entropy for β - power probability distribution and constant utility distribution.

## 3. Information Generating Function for Power Distributions

Let

(11) $\quad \Gamma_n = \{P = (p_1, p_2, \ldots, p_n, \ldots): p_i \geq 0, i \in N, \sum_{i \in N} p_i \leq 1\}, n = 2, 3, \ldots$

denote the set of all finite discrete ($n$ - ray) generalized probability distributions.

Consider the power distribution

(12) $\quad P^\beta = \left(\frac{p_i^\beta}{\sum_{i \in N} p_i^\beta}\right), \beta > 0$

obtained from (11). Replacing $p_i$ by $\frac{p_i^\beta}{\sum_{i \in N} p_i^\beta}$ in (2.1), we obtain

(13) $\quad I(P^\beta, U, t) = \sum_{i \in N} \left(\frac{p_i^\beta}{\sum_{i \in N} p_i^\beta}\right)^{1-u_i(1-t)}, t \geq 1$

which can be taken as the generalized information generating function for probability distributions defined by (11). From (13), for constant utility distribution, it is clear that

$$I(P^{(\beta)}, U, t) = I(P^\beta, U, t) \left(\sum_{i \in N} p_i^\beta\right)^{1-u(1-t)}$$

Here $P^\beta = \left\{\frac{p_i^\beta}{\sum_{i \in N} p_i^\beta}\right\}_{i \in N}$ and $P^{(\beta)} = \{p_i^\beta\}_{i \in N}$. The following figure shows the variation of new IGF given by (8) for uniform and non-uniform probability and utility distributions.

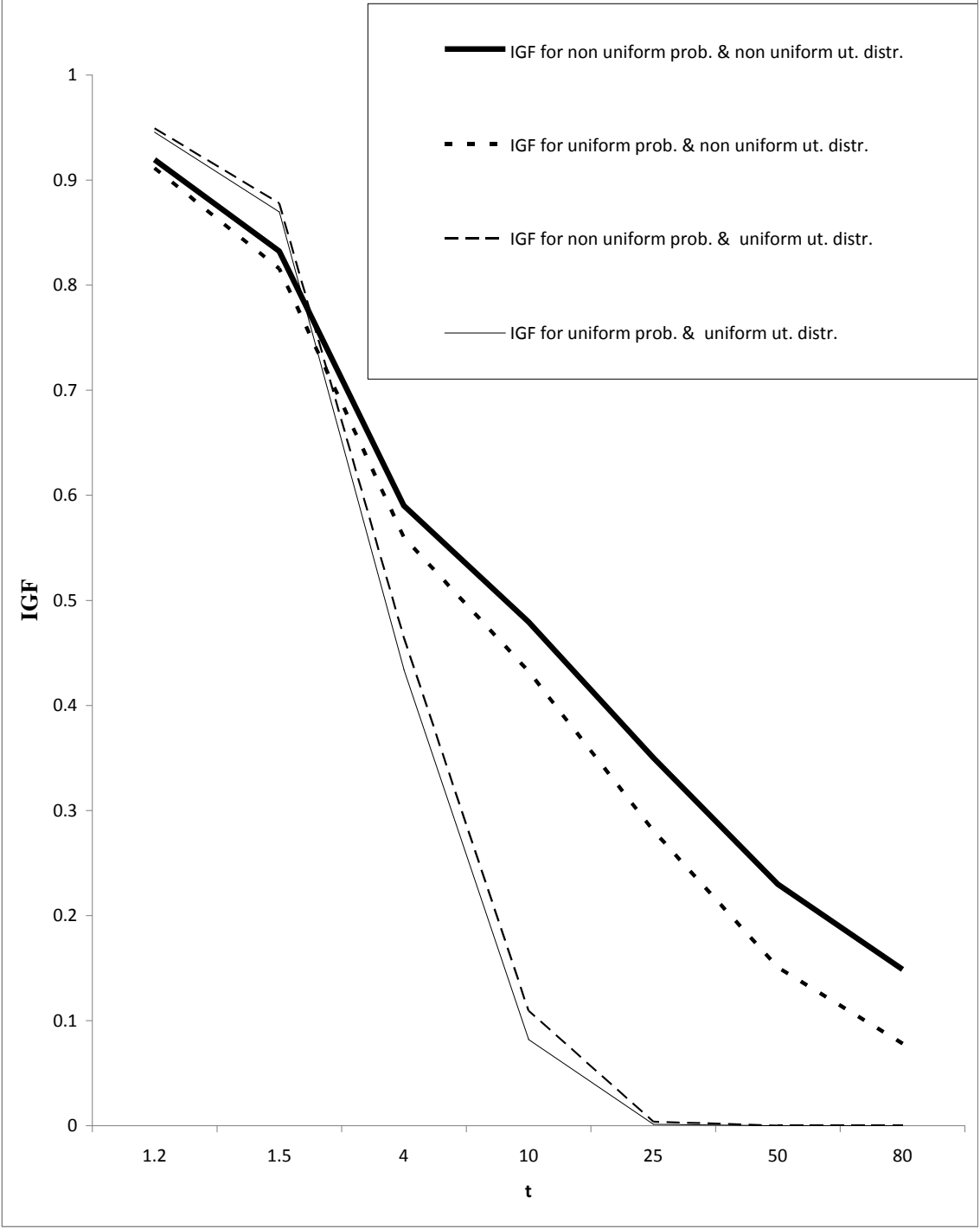